# Coupled Cluster Benchmark of New Density Functionals and of Domain Pair Natural Orbital Methods: Mechanisms of Hydroarylation and Oxidative Coupling Catalyzed by Ru(II) Chloride Carbonyls

Irena Efremenko[1] and Jan M.L. Martin[1, a)]

[1] *Department of Organic Chemistry, Weizmann Institute of Science, 76100 Reḥovot, Israel*

[a)]Corresponding author: gershom@weizmann.ac.il

**Abstract.** In the present work we tested the performance of several new functionals for studying the mechanisms of concurrent reaction of hydroarylation and oxidative coupling catalyzed by Ru(II) chloride carbonyls. We find that DLPNO-CCSD(T) is an acceptable substitute for full canonical CCSD(T) calculations; that the recent ωB97X-V and ωB97M-V functionals exhibit superior performance to commonly used DFT functionals; and that the revised DSD-PBEP86 double hybrid represents an improvement over the original, even though transition metals were not involved in its parametrization.

## INTRODUCTION

Synthetic methods allowing one-step C-C bond formation through homogenous-catalyst-mediated transformation of C-H bonds have become increasingly important in both industry and academia.[1,2] Potential routes for selective C-H bond activation and subsequent C-C bond formation in alkenes and aromatic compounds include hydroarylation by addition of aromatic C–H bonds across an unsaturated C=C bond, or an oxidative coupling that preserves the double bond. The latter reaction, while a highly desirable industrial goal, is challenging from the synthetic point of view. Pioneering examples of Ru-catalyzed coupling of aromatic carbon-hydrogen bonds with olefins were reported two decades ago.[3,4] Since then (for a review, see Ref. 2), an increasing number of examples catalyzed by Rh, Ru and Pd have been published, but the mechanistic aspects of the reactions have only been addressed by experimental methods.

Motivated by the experimental results of Milstein and coworkers,[4] over a decade ago we started to explore the mechanisms of the concurrent reactions of oxidative coupling and hydroarylation of methyl acrylate (MA) catalyzed by Ru carbonyl complexes (Scheme 1) using hybrid and double hybrid DFT families. These calculations showed that either proton elimination by chloride ion, or hydrogen transfer to coordinated olefin, can serve as the initial step of aromatic C-H bond activation, while oxidative addition mechanism could be excluded. Proton elimination proceeds via transition state **TS1** and initiates oxidative coupling with olefin (**TS2**) according to Scheme 2a. Next, interaction of Ru hydride intermediate **Int3** with MA and HCl regenerates the initial RuCl$_2$ carbonyl (**TS4a**). Alternatively, the catalytic cycle could be closed by interaction of **Int3** with MA and benzene, yielding **Int1** (**TS4b**). Hydrogen transfer to MA (**TS1a** and **TS1b**) causes coexistence of phenyl and one of the two isomeric alkyl ligands in the Ru coordination sphere (**Int4** and **Int5**). This mainly leads to the hydroarylation products via **TS3** and **TS3a** (Scheme 2,b), but coordination of the second olefin molecule followed by oxidative coupling is also possible.

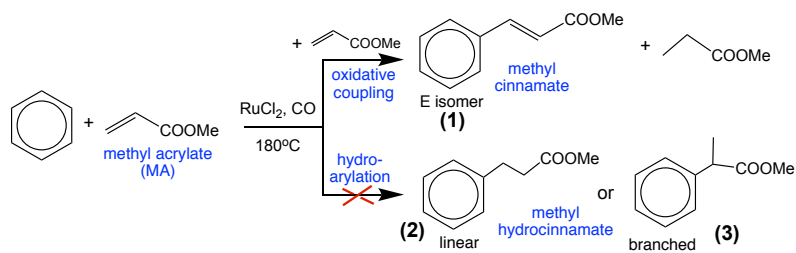

**Scheme 1.** *Possible interactions between MA and benzene in presence of Ru(II) chloride carbonyl complexes in benzene solution: oxidative coupling (observed) and hydroarylation (not observed).*

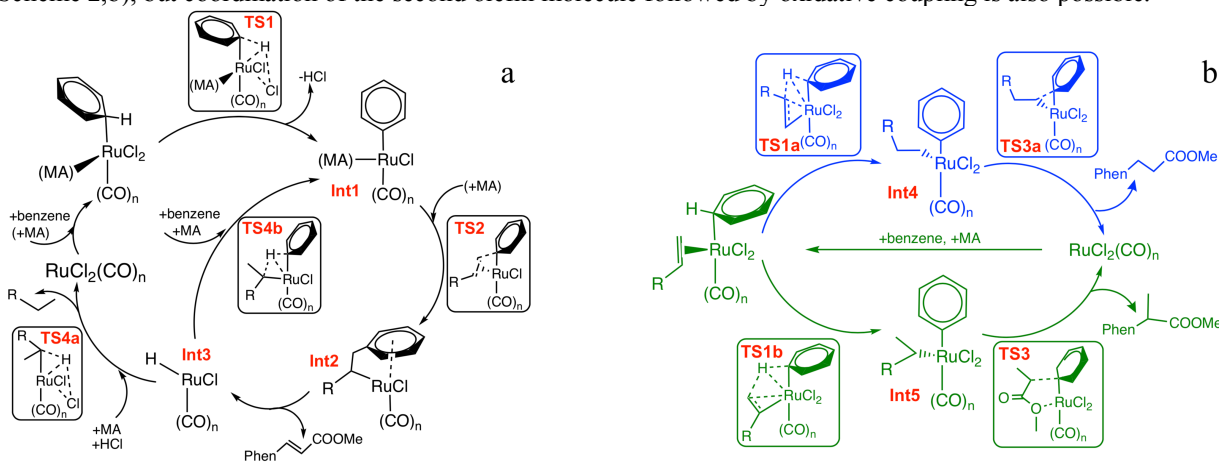

***Scheme 2.*** *Mechanisms of MA interactions with benzene in presence of Ru(II) chloride carbonyl complexes: oxidative coupling (a) and hydroarylation (b).*

We found that the activation barriers, the relative energies of the key intermediates, and the overall direction of the catalytic reaction strongly depend on the composition of the Ru coordination sphere. Ru complexes that could form in the reaction mixture, and serve as initial species of catalytic cycles, are shown in Scheme 3. In all the complexes, chloride anions are strongly bound to the metal atom; the only exception is $RuCl_2(CO)_4$(benzene) (rightmost complex in the 2nd row of Scheme 3) showing CO insertion into one of the Ru-Cl bonds. Moreover, the energetic results, particularly calculated activation barriers, differ between computational approaches and were hard to reconcile with experimental observations.

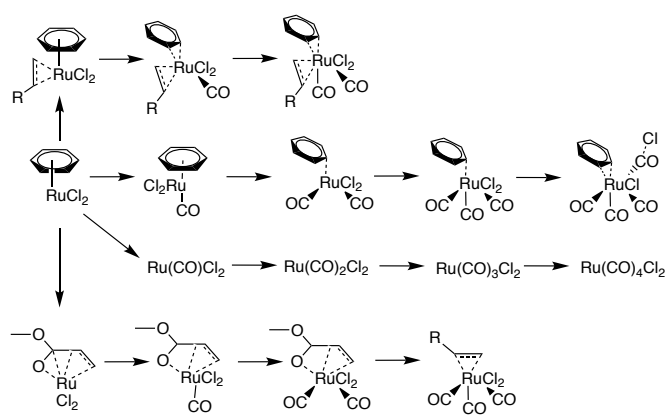

***Scheme 3.*** *Calculated $RuCl_2$ complexes with benzene, CO and MA. Arrows indicate that complex [$RuCl_2$(benzene)] is used as a reference.*

Therefore, for this specific case, we will attempt to obtain rigorous first-principles results by means of coupled cluster theory near the complete basis set limit, and use these results to assess the performance of more affordable computational methods. We believe our results have broader relevance for modeling mechanisms of catalytic reactions mediated by transition metal complexes.

The CCSD(T) method[5] is considered the "gold standard" of quantum chemistry. However, the computational cost of canonical CCSD(T) calculations scales as $O(N^7)$ and becomes prohibitively high for mechanistic studies of practical transition metal catalysis problems. Recently developed domain pair natural orbital methods, such as DLPNO-CCSD(T) of Neese and coworkers[6] and PNO-LCCSD(T) of Werner and coworkers,[7] scale almost linearly with system size (at least for closed-shell cases) and, in the main group, provide similar accuracy to the corresponding canonical calculation. Recently, benchmark studies of the performance of density functionals for transition metal problems, using DLPNO-CCSD(T) for calibration, have started appearing for reaction energies[8] and barrier heights.[9] Since the Ru complexes shown in Schemes 2 and 3 are still barely tractable by canonical methods, this enables us to assess the "domain error" for real-size transition metal complexes. In this paper, we assess both DFT and PNO methods against canonical CCSD(T) for the hydroarylation and oxidative coupling of benzene and methyl acrylate (MA) catalyzed by $RuCl_2$-carbonyl complexes, as a representative example for the complex mechanisms of homogeneous catalytic reactions.

Having in mind specific computational problems usually addressed in such mechanistic studies, the calculations were divided into four groups: (i) overall reaction energies (Scheme 1) that do not involve transition metals; (ii) relative energies of stable $RuCl_2$ complexes with CO, benzene and MA (Scheme 3) that should reproduce breaking and formation of the metal-ligand bonds and deep alterations of the coordination sphere and electronic structure of the metal atom; (ii) energies of key intermediates along reaction pathways catalyzed by different Ru complexes (Scheme 2) and (iv) barrier heights along these reaction pathways. It would be natural to calculate relative energies of the carbonyl complexes relative to $RuCl_2$, however, our calculations revealed that it has a triplet ground state, and that the singlet is essentially purely biradical (which is also reflected in the pathological $D_1$ diagnostic[10] value of 0.435). Discussion of the open-shell calculations are beyond the scope of this preliminary report; therefore, we used $RuCl_2$(benzene) as a reference, as $RuCl_2$ will anyway have no independent existence under the experimental conditions (benzene solvent). At all levels, 1:1 exchange of benzene with CO or MA is energetically unfavorable; all other complexes are exothermic with respect to the reference. The relative energies of key intermediates and transition states along each reaction path were calculated relative to the initial form of the catalyst, $(C_6H_6)(CO)_nRuCl_2$ (n=0-4).

## COMPUTATIONAL METHODS

The Weigend-Ahlrichs basis set family[11] def2-TZVP, def2-TZVPP, and def2-QZVPP was used throughout.

Reference geometries were optimized at the PBE0-D3BJ/def2-TZVP level[12,13] using Gaussian 09;[14] identities of transition states were verified by frequency and intrinsic reaction coordinate calculations.

At the final geometries, canonical CCSD(T)/def2-TZVPP single-point energy calculations were performed using MOLPRO 2018,[15] both using default frozen cores and including Ru(4s,4p) subvalence orbitals (which *are* correlated by default in ORCA. We found in this work that the mean absolute effects of Ru(4s,4p) subvalence correlation on carbonyl ligand energies and transition states are both a nontrivial 1.0 kcal/mol.) DLPNO-CCSD(T)[16] and the version with improved iterative triples, DLPNO-CCSD(T1),[17] were calculated with the def-TZVPP basis set using ORCA,[18] likewise DLPNO-CCSD(T)/def2-QZVPP calculations were done for basis set extrapolation using the simple $L^{-3}$ formula.[19] TightPNO cutoffs[20] were used to reduce domain discretization error; we found in the present work that DefaultPNO causes errors up to 3.5 kcal/mol in energy differences, and hence do not recommend its use.

In addition, single-point DFT calculations with a number of DFT functionals were carried out using ORCA. Aside from PBE0 already mentioned, these include: (a) the Berkeley "combinatorially optimized"[21] B97M-V, ωB97X-V and ωB97M-V; (b) the M06 family:[22] M06-L, M06 and M06-2X; (c) TPSS[23] and two different hybrids thereof, namely, TPSSh and TPSS0 (10% and 25% HF exchange, respectively); (d) both the original double-hybrid DSD-PBEP86-D3BJ[24] and its reparametrized version revDSD-PBEP86-D4;[25] in the latter, D3BJ also replaced with the very recently published next-generation D4 model.[26] As basis set convergence of double hybrids tends to be dominated by the MP2-like term, we carried out def2-TZVPP and def2-QZVPP calculations and applied $L^{-3}$ basis set extrapolation;[27] for the remaining DFT functionals we applied def2-TZVPP except for ωB97, which were accurate enough that we also tried def2-QZVPP. (Changes are on the order of 1 kcal/mol.) GRID6 was used in all DFT calculations.

In all Orca calculations, the RIJCOSX approximation[28] was employed, as well as the RI-MP2 approximation[29] for the double hybrids, in conjunction with the respective appropriate auxiliary basis sets[30,31] for the def2 family.

## RESULTS AND DISCUSSION

MAD (mean absolute deviation) and RMSD (root mean square deviation) error statistics with respect to our best CCSD(T) basis set limit estimates are shown in Figure 1 for the four types of energy differences. As expected, the smallest deviations were found for the overall reaction energies, which involve only main group elements. The hybrid functionals of the M06 family (MAD=0.68 and 0.57 kcal/mol for M06 and M06-2X, respectively) and double hybrid revDSD-PBEP86 functionals (MAD=0.64) show the best performance in this group. However, overall for the four criteria, the best results were obtained using ωB97M-V and ωB97X-V range-separated hybrids as well as by the revDSD-PBEP86 double hybrid, with accuracy between DLPNO-CCSD and DLPNO-CCSD(T). On the 3$^{rd}$ rung (meta-GGA) of the Jacob's Ladder, B97M-V performed best for reaction energies and carbonyl complex stabilities; however, M06-L showed similar performance, and TPSS outperformed them for barrier heights (MAD 3.69 vs. 4.54 for M06-L and 5.47 for B97M-V). The revDSD-PBEP86 functional outperforms the original DSD-PBEP86 for all four groups of calculations, whereas DSD-SCAN-D4 improved on DSD-PBEP86-D3BJ only for the carbonyl complexes. The DLNPO-CCSD(T) approach shows very close agreement (MAD=0.35 kcal/mol) with canonical CCSD(T) for the reaction energies, while for other groups there is a bit more daylight between them (MAD=1.04 for carbonyls, 1.56 for intermediates, and 1.58 kcal/mol for TSes). Using the DLNPO-CCSD(T1) approach with improved perturbative triples decreases these statistics to 0.60, 0.80, and 1.02 kcal/mol, respectively. The RMSD/MAD ratios are close to the theoretical value[32] (for a normal distribution) of $\sqrt{(\pi/2)} \approx 1.2533$, for carbonyls and intermediates but much larger for TSes, indicating an outlier (RMSD=0.80, 0.86, 2.05 kcal/mol): we note that TS2-CO2 is essentially biradical (and has $D_1$>0.4). If we exclude it, MAD and RMSD drop to quite pleasing values of 0.63 and 0.73 kcal/mol, respectively.

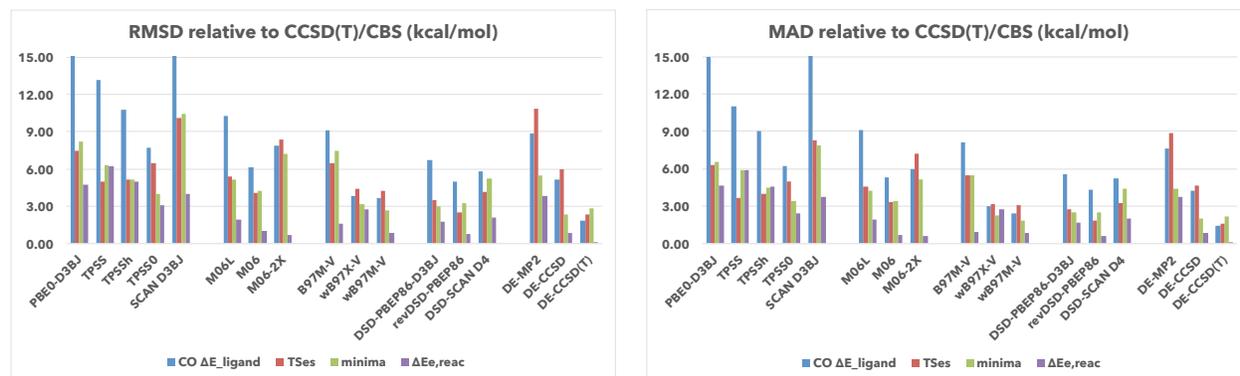

**Figure 1.** RMSD and MAD relative to CCSD(T)/CBS for the four types of energetics considered.

At all levels except for DLPNO-MP2, the highest RMSD and MAD values were found for the first reaction, i.e. dissociation of benzene and association of CO and MA ligands. Most density functionals tend to overbind the ligands; detailed analysis shows that for one group of functionals (PBE0-D3BJ, SCAN-D3BJ, TPSS, M06L and DSD-PBEP86 family, Fig. 2a) the error becomes greater with increasing number of CO ligands; the other group (M06, M06-2X, TPSS0, TPSSh, and B97, Fig. 2b) exhibits a less pronounced opposite trend, especially in the presence of coordinated benzene.

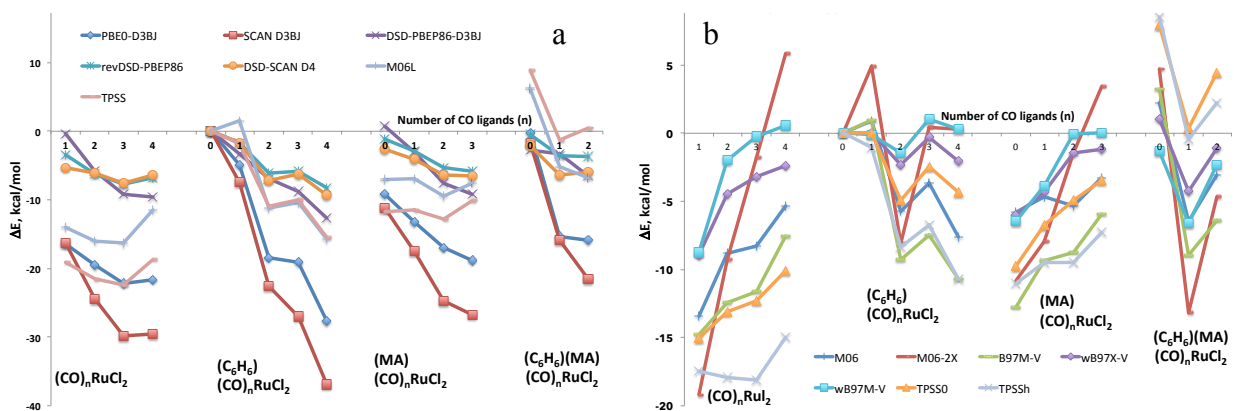

**Figure 2.** Energy deviation relative to CCSD(T)/CBS in different $RuCl_2 (CO)_n$ complexes as a function of n.

Our findings are consistent (see also our companion paper in the present volume) with the findings of Najibi and Goerigk[33] and ourselves[25] for the very large GMTKN55 main-group benchmark[34] and of Iron and Janes[9] for the MOBH35 transition metal reaction benchmark: Notably, that the range-separated hybrids ωB97X-V and ωB97M-V acquit themselves particularly well, that revDSD represents an improvement over the original DSD not just for the main group but also transition metals, and that unlike for the main group where empirical double hybrids are clearly superior, they offer no clear advantage over ωB97M-V for transition metal reactions. Unlike Iron and Janes, however, who found the new DSD-SCAN double hybrid[25] to be among the best performers for MOBH35, we find it to be inferior to revDSD-PBEP86 and the ωB97$n$-V family for the present problem (Figure 1).

## ACKNOWLEDGMENTS

This research was supported by the Israel Science Foundation (grant 1358/15), the Minerva Foundation, and the Helen and Martin Kimmel Center for Molecular Design (Weizmann Institute of Science).